# Dipolar relaxation, conductivity, and polar order in AgCN

P. Lunkenheimer,[1,a)] A. Loidl,[1] and G. P. Johari[2]

[1] Experimental Physics V, Center for Electronic Correlations and Magnetism, University of Augsburg, 86159 Augsburg, Germany

[2] Department of Materials Science and Engineering, McMaster University, Hamilton, Canada ON L8S 4L7

a) Author to whom correspondence should be addressed: peter.lunkenheimer@physik.uni-augsburg.de

**ABSTRACT**

By using dielectric spectroscopy in a broad range of temperatures and frequencies, we have investigated dipolar relaxations, the dc conductivity, and the possible occurrence of polar order in AgCN. Conductivity contributions dominate the dielectric response at elevated temperatures and low frequencies, most likely arising from the mobility of the small silver ions. In addition, we observe dipolar relaxation dynamics of the dumbbell-shaped $CN^-$ ions, whose temperature dependence follows Arrhenius behavior with a hindering barrier of 0.59 eV. It correlates well with a systematic development of the relaxation dynamics with the cation radius, previously observed in various alkali cyanides. By comparison with the latter, we conclude that AgCN does not exhibit a plastic high-temperature phase with a free rotation of the cyanide ions. Instead, our results indicate that a phase with quadrupolar order, revealing dipolar head-to-tail disorder of the $CN^-$ ions, exists at elevated temperatures up to the decomposition temperature, which crosses over to long-range polar order of the CN dipole moments below about 475 K. Dipole ordering was also reported for NaCN and KCN and a comparison with these systems suggests a critical relaxation rate of $10^5$ - $10^7$ Hz marking the onset of dipolar order in the cyanides. The detected relaxation dynamics in this order-disorder type polar state points to glasslike freezing below about 195 K of a fraction of non-ordered CN dipoles.

## I. INTRODUCTION

The cyanides, *A*CN, can be regarded as a class of "diatomic" ionic crystals with a spherical $A^+$ cation and a dumbbell-shaped rigid $CN^-$ anion. At high temperatures, in many of these systems the $CN^−$ ions were found to undergo almost free rotation and, hence, establish a highly symmetric cubic phase.[1,2,3] The latter can be termed plastic phase, in analogy to the so-called plastic crystals, which consist of molecules whose centers of gravity are located on crystalline lattice positions while being orientationally disordered.[4,5,6] On cooling, *A*CN crystals are known to exhibit phase transitions into orientationally ordered phases, where in many cases quadrupolar order (orientational ordering of the dumbbell-axes; also termed elastic order[2]) is followed by dipolar order (additional order with respect to head and tail of the dipoles).[1,2] However, in some systems dipolar order is avoided and instead a glasslike state is formed, e.g., lacking head-to-tail order or revealing other types of dipolar disorder at low temperatures.[1,2,7,8] These states can be regarded as "orientational glass",[7] analogous to the disordered low-temperature states known to arise in mixed systems like KBr-KCN.[1,9,10] It is the general belief that the quadrupolar, elastic order is driven and mediated by lattice strains while dipolar order results from Coulombic interaction forces coupled to phonon excitations.[1,2] In general, orientationally disordered crystals were intensively studied in the past: It was argued that they are molecularly analogues to spin glasses[11] and represent model systems to study glass-transition phenomena corresponding to those observed in canonical supercooled liquids.[1,5,10] Moreover, they have high entropy release at their order-disorder transitions, which may be relevant for future cooling technology,[12] and they can exhibit high ionic conductivity, making them promising candidates as electrolytes, employed, e.g., in solid-state batteries.[13,14,15]

Studies of the alkali cyanides $A$CN with $A$ = Na, K, Rb, and Cs are numerous and KCN is the most investigated of them. Possible hints on its orientational order already started 100 years ago by a controversy based on the analysis of early x-ray studies: Bozorth[16] claimed reasonable intensity agreement by orienting the CN molecules along the body diagonals of the cubic cell, while Cooper[17] disagreed with this analysis and suggested a free-rotation model of the CN dumbbells. That a dynamic reorientational model for the plastic high-temperature phase of KCN is indeed correct was established in detail by Elliott and Hastings[18] and Price et al.,[19] both utilizing neutron-diffraction experiments. By a combination of neutron diffraction and a total-scattering study using the reverse Monte-Carlo Method, the temperature dependence of the orientational distribution was determined by Cai et al.[20] In the last decades, potassium cyanide turned out to be a model system in many aspects: A sequence of quadrupolar and dipolar order was identified by thermodynamic[21] and by neutron-diffraction experiments,[22] the latter revealing antiferroelectric ordering. Haussühl[23] reported the extreme softening of the shear elastic constant, constituting the plastic character of the high-temperature phase.[24] Michel and Rowe[25] investigated the rotation-translation coupling of acoustic and librational modes in full detail, showing how quadrupolar interactions are mediated via lattice strains. Finally, the dilution of the CN$^-$ ions in KCN by halogenide ions suppresses the onset of long-range dipolar order and drives the system into an orientational glass state.[26] Just as KCN, pure NaCN also exhibits a transition into a dipole-ordered, antiferroelectric state.[22,27] In marked contrast, such a transition is absent in RbCN and CsCN, resulting in frozen-in dipolar disorder at low temperatures.[1,2]

While studies of the alkali cyanides are numerous, published information on cyanides with the group-11 elements Cu, Ag, and Au is scarce. For CuCN, structural investigations revealed that this material is polymorphic at room temperature.[7,28] In any case, it exhibits a linear-chain arrangement of the CN ions, alternating with Cu ions along the $c$ axis with significant head-to-tail disorder.[7,29,30] Moreover, disorder due to random chain displacement along the $c$-axis was reported.[31] Finally, temperature-dependent specific-heat measurements pointed to a glasslike transition at about 360 K in this material which can be ascribed to the glassy freezing of the CN dipole orientations.[7,8] AuCN reveals a similar chainlike arrangement of the CN ions as in the Cu compound and head-to-tail disorder and random chain displacements were reported for this cyanide, too.[31,32,33]

Focusing on silver cyanide investigated in the present work, it is a colorless powder which decomposes at about 600 K. West had reported the crystal structure of AgCN[34,35] and obtained lattice constants of $a$ = 5.99 Å and $c$ = 5.26 Å at ambient temperature. He described its structure to be of rhombohedral symmetry and to be made of -Ag-C-N-Ag- chains along the crystallographic $c$ direction. The Ag-Ag distance within the chain corresponds to the lattice constant of 5.26 Å and the distance between the chains is 3.47 Å. Bowmaker et al.[32] and Hibble et al.[36] performed neutron powder-diffraction studies documenting a $R\bar{3}m$ symmetry. As a structural synopsis, these authors documented that in this cyanide the silver and cyanide ions form infinite linear chains being arranged parallel along $c$, with alternating displacements of neighboring chains, and with some experimental evidence that the dipoles are disordered with respect to head and tail. From total neutron diffraction, it was concluded that the Ag-C and Ag-N bond lengths are almost identical and equal to 2.06 Å.[36] To unravel the possible head-to-tail order of the CN dumbbells, Bryce and Wasylishen[37] undertook solid-state NMR experiments. They concluded that, at room temperature, for 30 ±10% of the silver sites the CN arrangement is -NC-Ag-CN- or -CN-Ag-NC-, which they termed "disordered". Furthermore, for 70 ±10% of the silver sites they found -NC-Ag-NC-, termed "ordered".[37] Putting these results into a language of possible polar order of the CN molecules carrying a permanent dipole moment, 100% head-to-tail disorder corresponds to antiferroelectric and 100% order to ferroelectric order.

Hence, at room temperature 70% of the CN molecules seem to be arranged in a ferroelectric fashion. While it seems clear that AgCN below melting exhibits quadrupolar order, i.e., a strict orientational order of the long CN dumbbell axes along the crystallographic $c$ direction, the question of dipolar or head-to-tail order of the CN molecules remains to be solved. Hence, we decided to perform dielectric experiments to identify possible polar order and dipole dynamics in the title compound.

## II. EXPERIMENTAL DETAILS

Silver cyanide powder with 99.9 purity was purchased from Aldrich and pressed into pellets of 0.3 - 0.5 mm thickness and 5.5 - 12.3 mm diameter. For temperature variation, we used a $N_2$-gas flow cryostat (Novocontrol Quatro). Before the measurements, all samples were kept at 120 °C in nitrogen atmosphere for several hours to remove any moisture. To perform broadband dielectric spectroscopy, silver paint was applied on opposite sides of the disk-shaped samples. A combination of several techniques was used:[38] At frequencies from 25 mHz up to 1 MHz, a frequency-response analyzer (Novocontrol Alpha-A analyzer) and an LCR meter (HP 4284A) were employed. Measurements from 1 MHz to about 1 GHz were performed with an impedance analyzer (HP 4291A) using a coaxial reflectometric setup.[39] This method is least affected by stray capacitance or other parasitic effects. To achieve a good match between the spectra at high and low frequencies, partly the low-frequency spectra were shifted, applying the same scaling for all temperatures of a measurement run. Measurements were performed both upon heating and cooling, revealing small irreversible changes of the dielectric properties when heating the sample above about 430 K, which may point to the onset of sample degradation. Therefore, we used the spectra taken upon heating for analysis and one should be aware that the data at the highest temperatures may have some uncertainty.

## III. EXPERIMENTAL RESULTS AND DISCUSSION

Figure 1 presents the dielectric constant $\varepsilon'$ (a) and the dielectric loss $\varepsilon''$ (b) (the real and imaginary part of the permittivity, respectively) of AgCN as function of frequency on double-logarithmic scales as measured for a series of temperatures between 225 and 470 K. The frequency regime extends from 25 mHz up to 1 GHz, depending on the devices employed for measurements at different temperatures. In Fig. 1(a), at elevated temperatures $\varepsilon'(\nu)$ strongly increases with decreasing frequency and finally reaches values beyond $10^3$ [not covered in Fig. 1(a)]. Moreover, Fig. 1(b) reveals a dramatic increase of the loss towards low frequencies. Both phenomena clearly signal the importance of contributions from conductivity and from electrode polarization (sometimes termed "blocking electrodes"), typical for ionic conductors.[40] Notably, such significant charge-transport effects, strongly superimposing the dipolar relaxation features (treated below), are absent in the permittivity spectra of the $A$CN cyanides with $A$ = Na, K, Rb, and Cs, shown in Ref. 2. It is well known that many crystalline silver compounds like AgI or $AgRb_4I_5$ exhibit considerable ionic conductivity.[41,42] Therefore, it seems reasonable to ascribe the conductivity in AgCN to the mobility of the relatively small $Ag^+$ ions. Earlier pressure-dependent conductivity measurements of AgCN also pointed to ionic charge transport at ambient pressure.[43]

The dielectric spectra of the cyanides with $A$ = Na, K, Rb, and Cs are known to exhibit well-pronounced relaxation phenomena, i.e., a steplike decrease of $\varepsilon'(\nu)$ with increasing frequency and peaks in $\varepsilon'(\nu)$, which arise from the reorientational CN motions.[2] In the AgCN spectra of Fig. 1, such relaxation phenomena are also visible, showing up in the real part at high temperatures and being most obvious in the imaginary part in a temperature regime between 250 and 424 K. However, in AgCN they are superimposed by the mentioned charge-transport contributions. These relaxation features shift to higher frequencies with increasing temperatures, indicating a thermally activated acceleration of the corresponding dynamics. Based on the results for the other dielectrically investigated *A*CN cyanides,[1,2] we assume that these relaxations arise from 180° flips of the CN molecules in their local potential, carrying a static dipole moment of order 0.3 D.[44] In the frequency and temperature regime covered by Fig. 1, these relaxation phenomena significantly loose intensity upon cooling. As the dipole moment of the CN dumbbells should be temperature independent, this finding signals a decreasing number of molecules participating in the relaxational process, which can be ascribed to polar ordering effects. Similar behavior in the dipolar ordered phases of KCN and NaCN was stated to reflect "the gradual disappearance of alignable dipoles due to the onset of a second-order phase transition into the antiferroelectric ordered state".[2] Within this framework, the relaxational dynamics observed below the transition temperature arises from dipoles that do not participate in the antipolar order. Such dipoles should be most numerous just below the transition and their number should continuously decrease upon cooling, in accord with the continuously growing order parameter of a second-order phase transition. In AgCN, below about 250 K the amplitudes of the relaxation phenomena are so weak that they are practically unobservable in the real part of the permittivity [Fig. 1(a)], partly because these contributions are hidden underneath $\varepsilon_\infty$, arising from high-frequency excitations.

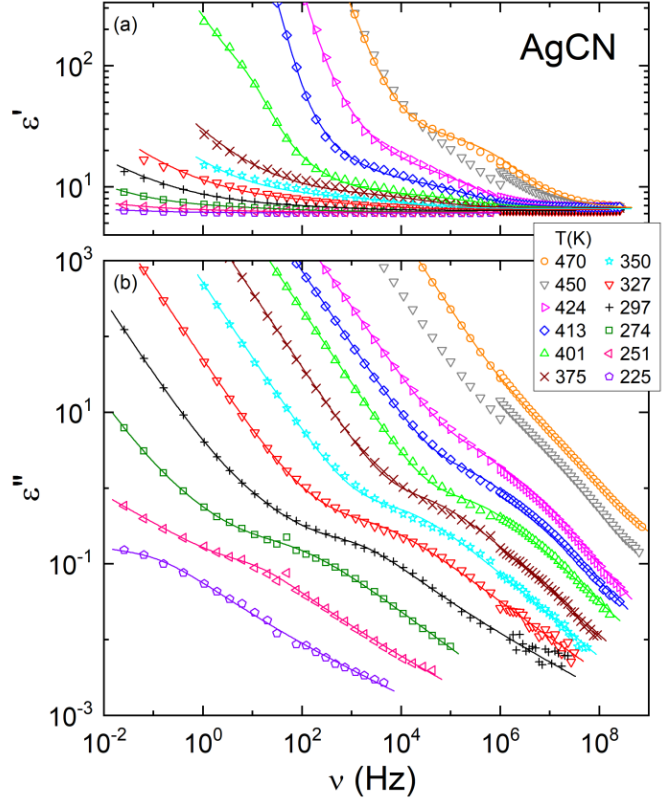

**FIG. 1.** Complex permittivity spectra of AgCN shown on double logarithmic scales for temperatures between 225 and 470 K. (a) Dielectric constant $\varepsilon'$ and (b) dielectric loss $\varepsilon''$. At 450 K, low- and high-frequency results do not perfectly match due to irreversibilities occurring at high temperatures. The lines are fits, which were simultaneously performed for the real and imaginary part, including a Cole-Cole function for the dipolar relaxation features, conductivity contributions, and (at $T \geq 401$ K) contributions from electrode polarization modeled by an RC equivalent circuit[40] (see text for details). Cole-Cole function gave the best fit to both the $\varepsilon'$ and $\varepsilon''$ spectra.

To gather quantitative information about the temperature dependences of the dc conductivity, the relaxation time $\tau$, and the relaxation strength $\Delta\varepsilon$, the complex dielectric spectra of Fig. 1 were analyzed assuming a dipolar relaxation process and additional ac- and dc-conductivity contributions to model the intrinsic sample properties. The overall intrinsic complex permittivity $\varepsilon^* = \varepsilon' - i\varepsilon''$, which is directly related to the complex conductivity $\sigma^* = \sigma' + i\sigma''$ via $\varepsilon^* = \sigma^*/(i\varepsilon_0\omega)$, then is given by:

$$\varepsilon^* = \varepsilon_\infty + \frac{\sigma_{dc}}{i\varepsilon_0\omega} + \frac{\sigma_{ac}^*}{i\varepsilon_0\omega} + \frac{\Delta\varepsilon}{1+(i\omega\tau)^{1-\alpha}} \tag{1}$$

Here $\omega = 2\pi\nu$ is the angular frequency and $\varepsilon_0$ the dielectric permittivity of free space. The quantity $\varepsilon_\infty$ results from high-frequency dipole-active excitations and contributes to the real part of the dielectric constant only. $\sigma_{dc}$ is the dc conductivity. The complex ac conductivity is given by the so-called universal dielectric response (UDR),[45] $\sigma_{ac}^* \propto (i\omega)^s$, with a frequency exponent $s \leq 1$. It is usually ascribed to hopping charge transport of localized charge carriers as found in various classes of disordered matter and in ionic conductors.[46,47,48] It will not be treated in detail here as we found only minor ac-conductivity contributions even at the highest temperatures, which helped improve the fits in the transition region from the conductivity-dominated to the relaxation-dominated frequency regime. The last term in Eq. (1) describes a Cole-Cole-type dipolar relaxation process.[49] The relaxation strength $\Delta\varepsilon$ corresponds to the difference of the limiting values of the real part of the permittivity for frequencies well below and above the relaxation frequency. $\tau$ is the mean relaxation time, and $\alpha$ is a width parameter leading to symmetrical broadening of the loss peaks compared to the Debye equation where $\alpha = 0$. The empirical Cole-Cole function was also used to describe the dielectric relaxation processes in the antiferroelectric cyanides NaCN and KCN.[2] For the spectra at $T \geq 401$ K, the intrinsic contributions to the permittivity, covered by Eq. (1), were insufficient to fit the experimental data, especially at low frequencies. To account for the above-mentioned significant electrode effects observed at these temperatures, we employed an equivalent-circuit approach as described in detail in refs. 40 and 50. There the thin insulating layers arising from the blocking of ion motion at the sample surfaces are modeled by a parallel RC circuit, connected in series to the bulk part of the sample.

The lines in Fig. 1 show fits with the above approach, simultaneously performed for the real and imaginary part of the permittivity. Overall, we arrived at a satisfying description of the frequency dependence of the dielectric permittivity of AgCN in a broad range of frequencies and temperatures. Due to the partly irreversible sample changes at high temperatures mentioned in section II, the low- and high-frequency spectra at 450 K do not match and, hence, we did not fit this data set. Considering the rather large number of fit parameters, especially at high temperatures where electrode effects had to be taken into account, we want to point out that the results on the most relevant parameters of the main relaxation feature ($\tau$, $\Delta\varepsilon$, and $\sigma_{dc}$) were robust against variations of the additional contributions, employed to fit the complete dielectric spectra of Fig. 1. The experimental permittivity spectra either reveal a well-defined shoulder in $\varepsilon''(\nu)$ or a clearly discernible step in $\varepsilon'(\nu)$ (or both), thereby helping to define these parameters.

In the following, we discuss the temperature dependence of the obtained fit parameters. Utilizing the simplest ansatz, assuming no or only weak cooperativity of the molecular reorientations, the mean relaxation time should exhibit an Arrhenius-type temperature dependence given by:

$$\tau = \tau_0 \exp\left(\frac{E_\tau}{k_B T}\right) \tag{2}$$

Here $k_B$ is the Boltzmann constant, $\tau_0$ represents a temperature-independent prefactor, corresponding to an inverse microscopic frequency (attempt frequency), and $E_\tau$ is the hindering barrier against dipolar reorientations (180° flips in the present case). In Fig. 2(a), we present an Arrhenius plot of the temperature-dependent relaxation time resulting from the fits in Fig. 1. The observed linear behavior indicates purely thermally activated Arrhenius-type temperature dependence [Eq. (2)], and from the slope of the linear fit (line) we deduce $E_\tau \approx 0.59$ eV. In light of the above discussion of the microscopic origin of the observed relaxation process, this is the energy barrier that the CN$^-$ ions, aligned along the crystallographic $c$ direction with two adjacent silver ions, must pass during their 180° reorientational motions. Arrhenius behavior of the orientational relaxation dynamics was also found for the $A$CN cyanides with $A$ = Na, K, Rb, Cs, and Cu.[2,8]

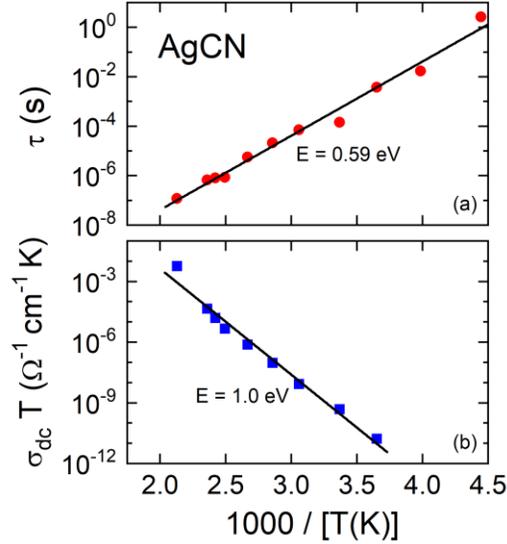

**Fig. 2.** Temperature dependence of the mean relaxation time $\tau$ and the dc conductivity $\sigma_{dc}$ of AgCN as obtained from the fits of the permittivity spectra shown in Fig. 1. (a) Arrhenius plot of $\tau$. (b) Temperature dependence of $\sigma_{dc}$ in a plot $\sigma_{dc}T$ vs. 1000/$T$, linearizing Eq. (3). The lines in (a) and (b) are fits with Eqs. (2) and (3), respectively, leading to energy barriers as indicated in the figures.

Figure 2(b) shows the temperature dependence of the dc conductivity obtained from the fits in Fig. 1. As discussed above, it presumably arises from the thermally activated long-range diffusion of silver ions through the crystalline lattice. For the temperature dependence of ionic dc conductivity, often a modified Arrhenius law is assumed:[51,52]

$$\sigma_{dc} = \frac{\sigma_0}{T} \exp\left(-\frac{E_{dc}}{k_B T}\right) \qquad (3)$$

Here $\sigma_0$ is a temperature-independent prefactor and $E_{dc}$ the hindering barrier for the ionic hopping processes. If Eq. (3) is valid for AgCN, the representation in Fig. 2(b) should linearize the experimental data, which indeed is the case. The deduced hindering barrier against silver diffusion is $E_{dc} \approx 1.0$ eV. As proposed long ago,[13] certain plastic crystals represent good ionic conductors, which may be ascribed to a paddle wheel or revolving door mechanism where the ionic mobility is enhanced by the reorientational dipole motions.[14,15,53,54] In strict sense, AgCN does not exhibit a plastic phase, as the CN molecules cannot rotate freely, but the CN dipoles still undergo 180° flips which even at ambient temperature are faster than ms [Fig. 2(a)]. Anyway, the finding of significantly different hindering barriers for dipolar reorientation and ionic diffusion (Fig. 2) points to the decoupling of both motions, making a paddle wheel or revolving door mechanism unlikely. It thus seems that the diffusive process of the silver ions is not directly influenced by the dipolar reorientations of the CN molecules and more likely occurs via defect paths or local structural distortions in a disordered crystalline lattice.

In general, polar order can be detected via the temperature dependence of the order parameter, which counts the normalized difference of ordered and disordered sites and is related to the relaxation strength $\Delta\varepsilon$. Figure 3 shows the temperature dependence of $\Delta\varepsilon$ deduced from the fits of the dielectric spectra (Fig. 1). As already revealed by the amplitudes of the spectra of Fig. 1, it continuously decreases upon cooling. This reduction of $\Delta\varepsilon(T)$ amounts about two orders of magnitude for temperatures between 470 and 225 K. As discussed above, it most likely signals the continuous decrease of the number of reorienting dipoles below a polar ordering transition. For order-disorder-type polar order, i.e., the ordering of dipoles that already exist above the transition temperature, a Curie-Weiss law, $\Delta\varepsilon \propto 1/(T_{c,p}-T)$ can be expected below the ordering temperature $T_{c,p}$.[55,56] In the inset of Fig. 3, we demonstrate that, at $T \geq 350$ K, the inverse of the relaxation strength is consistent with a linear decrease, indeed in accord with Curie-Weiss behavior. The extrapolated linear fit of the measured data (solid line in the inset of Fig. 3) crosses zero close to 475 K, which can be taken as an estimate of the critical temperature $T_{c,p}$ for polar ordering in AgCN.

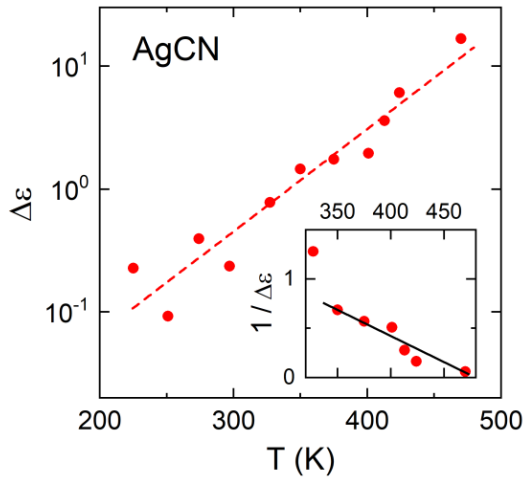

**Fig. 3.** Temperature dependence of the relaxation strength $\Delta\varepsilon$ of dipolar reorientations (semilogarithmic plot) as obtained from fits of the permittivity spectra (Fig. 1). The dashed line is a linear fit. The inset shows $1/\Delta\varepsilon(T)$ at high temperatures. The solid line indicates a linear fit crossing zero close to 475 K, which we consider as an estimate of the polar ordering temperature $T_{c,p}$.

Further parameters that determine the permittivity spectra of Fig. 1, but will not be discussed in full detail, are those defining the ac conductivity, the infinite dielectric constant, and the width of the relaxational peaks. As outlined earlier, ac conductivity in AgCN plays a minor role only. It was possible to fit the extra ac contributions with an exponent parameter $s$, decreasing from values of about 0.7 to 0.3 when temperature rises from 225 to 470 K. The infinite-frequency dielectric constant $\varepsilon_\infty$ is only weakly temperature dependent: Its moderate increase on increasing temperatures (from 6.1 to 6.8) is a well-known phenomenon in anharmonic crystals. Its absolute values compare reasonably with those observed in the alkali cyanides, which in the orientationally ordered phases vary between 5 and 8.[2] The width parameter $\alpha$, describing the symmetric loss-peak broadening of the Cole-Cole function, remains essentially temperature independent, with values scattering around 0.35 [note that $\alpha = 0$ describes monodispersive Debye relaxation, cf. Eq. (1)].

Such significant broadening, compared to the Debye case, is considered a hallmark feature of glassy freezing and usually ascribed to a distribution of relaxation times due to heterogeneity.[57,58] In the present case, the latter probably stems from a distribution of energy barriers around the CN molecules indicating relatively strong static disorder. For KCN, NaCN, and CsCN, smaller width parameters of the order of 0.1 - 0.2 were reported, slightly increasing on decreasing temperatures.[2] The enhanced loss-peak widths in AgCN signal a higher degree of disorder, which also could result from Ag diffusion processes as indicated by the relatively high dc conductivity.

## IV. COMPARISON WITH THE ALKALI CYANIDES

In the following, we will compare the relaxation dynamics of AgCN with those observed in the alkali cyanides, which all are well characterized by dielectric spectroscopy.[2,26,59,60,61,62,63] This comparison should also allow for some further conclusions concerning possible polar order in AgCN. As discussed in section I, the alkali cyanides $A$CN ($A$ = Na, K, Rb, and Cs) undergo structural phase transitions from high-temperature cubic phases (NaCl-type in Na, K, and Rb; CsCl-type in CsCN) into low-temperature phases with structural distortions and long-range quadrupolar order (i.e., with orientationally ordered dumbbell-axes). The phase transitions occur at 132, 168, 193, and 288 K for the Rb, K, Cs, and Na compounds, respectively.[26] At the onset of quadrupolar order, KCN and NaCN transform into orthorhombic phases,[22] while RbCN has a low-temperature monoclinic[64] and CsCN a trigonal structure.[65] Upon further cooling below these quadrupolar transitions, KCN and NaCN reveal an important difference compared to the Rb and Cs compounds: They exhibit an additional phase transition into antiferroelectric order below 83 K[66] and 172 K,[67] respectively, while no polar phase transitions were identified in RbCN and CsCN.[2,64] Instead, in the latter two the head-to-tail disorder freezes in at low temperatures and then both systems can be characterized as dipolar glasses, analogous to spin glasses. Despite these differences in their low-temperature states, the dipolar relaxation dynamics in all four compounds behave similar in many respects, indicating that, with the onset of long-range polar order in KCN and NaCN, the local energy potential surface sensed by the dipoles that do not take part in the polar order is not drastically changed.

Concerning the relaxation dynamics in the alkali cyanides, Lüty and Ortiz-Lopez[26] made an interesting and unexpected observation: They found that the quadrupolar ordering transition in the alkali cyanides is triggered by a critical dipolar relaxation rate $1/\tau_{c,q} \approx 2\times10^{10}$ Hz. On cooling, quadrupolar order sets in when molecular reorientations fall below this rate. Sethna and coworkers[68] rationalized the occurrence of such a critical relaxation rate within mean-field theory, explaining it by the fact that the energy barrier against dipolar reorientations depends on the local environment, set up by the quadrupolar interaction forces mediated by lattice-strain energies. The results of Lüty and Ortiz-Lopez[26] are reproduced in Fig. 4, where the temperature dependence of the mean relaxation rates $1/\tau$ of all alkali cyanides is indicated in an Arrhenius-type representation, analogous to Fig. 1 of that work (note the inverted $1/T$ axis). In Ref. 26, $\tau(T)$ of the cyanides was collected from experiments utilizing ionic thermal conductivity, dielectric spectroscopy, and NMR techniques. All these data could be well fitted by thermally activated behavior, Eq. (2), and for clarity reasons, in Fig. 4 we show the fit curves of these data only (solid lines)[2,26]. For completeness, we also include data reported by Knopp et al.[61] for CsCN (dashed line in Fig. 4), which were determined by dielectric spectroscopy only and are similar to those in Ref. 26. The increasing slopes when moving from RbCN ($E_\tau = 0.10$ eV) to NaCN ($E_\tau = 0.28$ eV)[2] imply increasing hindering barriers. The attempt frequencies for all compounds are of order $10^{14}$ - $10^{15}$, representing typical values of a microscopic excitation frequency.[2] In Fig. 4, the closed squares show the quadrupolar phase-transition temperatures $T_{c,q}$ of the alkali cyanides, where quadrupolar elastic order is established. The light-blue bar indicates the mentioned critical relaxation rate of about $1/\tau_{c,q} \approx 2\times10^{10}$ Hz, below which all alkali cyanides reveal quadrupolar order.[26]

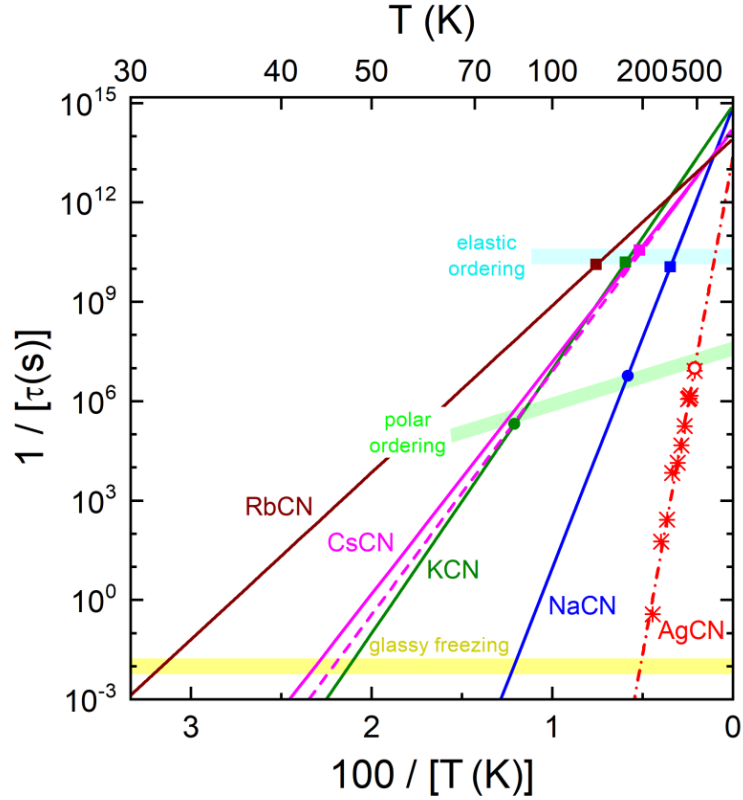

**Fig. 4.** Dipolar relaxation rates in AgCN compared to those observed in the alkali cyanides by Lüty and Ortiz-Lopez[26] (we use here the same Arrhenius representation as in Fig. 1 of Ref. 26; note the inverted x-axis with zero at the right). The data from Ref. 26 are represented by the corresponding Arrhenius fit lines (solid lines; parameters taken from Ref. 2). An alternative Arrhenius fit for CsCN as reported by Knopp et al.[61] is indicated as dashed line. The stars show the experimental results on AgCN as deduced in the present work and the dash-dotted line indicates the corresponding Arrhenius fit. The filled circles are the critical temperatures $T_{c,p}$ marking the onset of antiferroelectric order in KCN[66] and NaCN.[67] For AgCN, the open circle indicates the suggested polar ordering temperature $T_{c,p} \approx 475$ K, estimated from the Curie-Weiss fit of $\Delta\varepsilon(T)$ (inset of Fig. 3). The green bar connects the relaxation rates at $T_{c,p}$. The closed squares represent the elastic, quadrupolar ordering temperatures $T_{c,q}$.[26] The blue bar indicates the critical dipolar relaxation rate $1/\tau_{c,q} \approx 2\times10^{10}$ Hz as determined in Ref. 26. The yellow bar corresponds to the region of glassy freezing, where the molecular dynamics has a relaxation time of about 100 s.

Before discussing our AgCN results in the framework of these correlations, we draw attention to a striking finding in Ref. 26: In the complete temperature and frequency regime documented in Fig. 4, all alkali cyanides exhibit pure Arrhenius laws without any indications of anomalies due to critical dynamics. Note that all systems undergo elastic, quadrupolar order (transition temperatures $T_{c,q}$ indicated by the closed squares in Fig. 4), but NaCN and KCN exhibit additional long-range antiferroelectric order (closed circles), while RbCN and CsCN remain disordered (no head-to-tail order) down to the lowest temperatures. The polar ordering transitions in KCN and NaCN represent classical examples of order-disorder type phase transitions.[69] In order-disorder type polar materials the temperature dependence of the mean relaxation time $\tau$ should exhibit critical divergence close to the transition.[55,56] However, this often is superimposed by thermally activated behavior.[69,70,71,72] The dominance of the latter and the absence of any critical behavior of the relaxation rates of KCN and NaCN then may be explained by the fact that the critical enhancement is limited to a small temperature regime around $T_{c,p}$, being overlooked in canonical dielectric experiments.

Looking into the original data of Ortiz-Lopez et al.,[2] clear anomalies were observed in the temperature dependence of the real part of the dielectric constant. However, to determine mean relaxation rates, frequency-dependent measurements were performed in temperature steps of about 5 K, in KCN just reaching the onset of antiferroelectric order and in NaCN not covering the critical region. Just as for the alkali cyanides, no indication of a critical enhancement of the relaxation rate is detected in the present results on AgCN, where measurements could only be performed below the polar transition temperature, due to the irreversibility of the measured values at high temperatures, mentioned in section II.

When comparing the temperature-dependent relaxation times of AgCN (stars and dash-dotted line) to those observed in the alkali cyanides (solid lines), Fig. 4 documents that the dipolar dynamics of all these materials correlates very well: All compounds show pure Arrhenius behavior with attempt frequencies of order $10^{13}$ - $10^{15}$ s (read off at $100/T = 0$) and with energy barriers that tend to continuously increase with decreasing cation radius. Interestingly, this also is the case for CuCN, for which an energy barrier of 0.98 eV was deduced from calorimetric experiments[8] (larger than $E_\tau \approx 0.59$ eV of AgCN) and whose $Cu^+$ ion is even smaller than $Ag^+$. Only CsCN deviates from this trend, which may be related to its different structure, both in the plastic and the quadrupolar-ordered phases.

The yellow bar in Fig. 4 indicates the region of glassy freezing of the orientational dynamics, formally defined by a relaxation time $\tau(T_g^o) \approx 100$ s at an orientational glass-transition temperature $T_g^o$. However, one should note that only CsCN and RbCN undergo an orientational glass transition in the original sense, where all dipoles freeze into a disordered arrangement. In contrast, KCN and NaCN instead reveal long-range head-to-tail order and only part of the dipoles can exhibit such freezing. As discussed above and suggested in Ref. 2, within the scenario for order-disorder polar phase transitions, the detected relaxation dynamics should only arise from those dipoles that are not participating in the polar order. While their number continuously diminishes upon cooling, nevertheless their relaxation time also increases beyond 100 s, just as in the Rb and Cs compounds. This implies that their reorientational mobility essentially freezes in upon further cooling. Indeed, weak but significant glass-transition anomalies were detected by measurements of thermally stimulated depolarization currents in these systems.[2] Interestingly, this glasslike freezing of the remaining "free" CN molecules should lead to a certain degree of essentially static orientational disorder in KCN and NaCN at low temperatures, which prevents the realization of complete polar order of all dipoles at lowest temperatures in these systems. As the present work provides evidence for a polar ordered state in AgCN, featuring significant dipolar dynamics, this scenario is also valid for that cyanide. From the Arrhenius fit shown in Fig. 4, we obtain a glasslike freezing temperature of about 195 K. Edgar F. Westrum, Jr. in 1998 had informed one of the authors (GPJ) that adiabatic calorimetry performed by his group has found a glass-like transition endotherm in the temperature range 250-275 K in both normal and sintered AgCN. This is a significant discrepancy to $T_g^o \approx$ 195 K determined in the present work. The latter is mainly based on the relaxation times determined from the low-frequency spectra where $\Delta\varepsilon$ is small and no step in $\varepsilon'(\nu)$ is observed [Fig. 1(a)]. However, at least at 250 K, a clear shoulder shows up in the dielectric-loss spectrum, far beyond experimental uncertainty [Fig. 1(b)]. Its analysis leads to a relaxation time of about 17 ms [Fig. 2(a)], much shorter than $\tau \approx 100$ s, expected at the glass transition. Therefore, this discrepancy remains unresolved. Further measurements are necessary to clarify the origin of this discrepancy, e.g., neutron-scattering experiments determining the temperature dependence of the Debye-Waller factor which should show an anomaly at the glass transition.[73]

Despite significant structural differences, we assume that the critical relaxation rate $1/\tau_{c,q} \approx 2\times10^{10}$ Hz reported for the alkali cyanides[26] also determines the quadrupolar ordering temperature $T_{c,q}$ of AgCN. However, when extrapolating the fit line of $\tau(1/T)$ of AgCN shown in Fig. 4 to this critical rate (blue bar), this leads to a value $T_{c,q} \approx 970$ K, well above the decomposition temperature of about 600 K. Hence, most likely AgCN does not exhibit a high-temperature plastic phase characterized by freely rotating CN dumbbells, as observed in the alkali cyanides. Instead, the quadruple-ordered phase, where 180° dipolar flips of the CN molecules should still be possible, should prevail up to highest temperatures. As suggested by the decrease of $\Delta\varepsilon(T)$ discussed in section III (Fig. 3), these flips should start to cease below the polar ordering transition at about 475 K. Here we propose, in analogy to a critical relaxation rate at the quadrupolar phase transition, a critical relaxation rate for polar ordering. In Fig. 4, the antiferroelectric transition temperatures in KCN and NaCN are indicated by full circles. They suggest a slightly increasing critical relaxation rate for dipolar order with decreasing ionic radius (green bar in Fig. 4). This is well consistent with $T_{c,p} \approx 475$ K (open circle), estimated for AgCN from the temperature dependence of the relaxation strength $\Delta\varepsilon$ (inset of Fig. 3).

As discussed above, $\Delta\varepsilon(T)$ of AgCN continuously decreases from the highest temperatures measured, as expected below a polar order-disorder phase transition (Fig. 3).[55,56] In KCN and NaCN, such a decrease was also observed.[2,62,63] For AgCN, we find that, at high temperatures, this decrease is consistent with Curie-Weiss behavior with a critical temperature around 475 K (inset of Fig. 3). While this points to polar order of AgCN below ~ 475 K, the question remains whether it is of ferroelectric or antiferroelectric nature. We mentioned earlier that the polar ground state in the related alkali cyanides, NaCN and KCN is antiferroelectric. Notably, when assuming pure Coulombic interaction forces, the ground state in the two compounds would be ferroelectric, and the observed antiferroelectric ground state results from models taking electric-dipole dressing effects into account, which include cationic displacements.[74,75] AgCN is a linear-chain compound and ferroelectric order of the dipoles within the infinite Ag-CN chains seems to be the canonical ground state. Assuming pure Coulombic interactions alone, neighboring chains would exhibit antiferroelectric order. However, taking the results from NMR experiment on AgCN serious, which document that 70% of the CN dipoles are ordered in a ferroelectric fashion at room temperature,[37] clearly favors a ferroelectric ground state.

## V. SUMMARY AND CONCLUSIONS

In summary, we have performed a detailed dielectric investigation of the conductivity and dipolar relaxation dynamics of AgCN in a broad frequency and temperature range. The obtained permittivity spectra reveal a superposition of relaxational and conductivity contributions. The latter point to significant charge transport which we find to be thermally activated with an energy barrier of 1.0 eV. It can be ascribed to the mobility of the rather small $Ag^+$ ions and is decoupled from the detected relaxation dynamics arising from the reorienting dipolar CN dumbbells. The relaxation strength of the latter strongly decreases with decreasing temperature and is consistent with polar order of order-disorder type, forming below $T_{c,p} \approx 475$ K. In light of the findings from NMR measurements in Ref. 37, this order can be assumed to be of ferroelectric nature.

The dipolar relaxation rate follows Arrhenius temperature dependence. Comparing its absolute values and temperature development with the relaxation rates of the previously investigated alkali cyanides[2,26] and of CuCN[8] reveals a systematic development in dependence of the cation radius. The application of the critical-relaxation-rate scenario, proposed for the alkali cyanides,[26] indicates that AgCN does not have a plastic high-temperature phase with isotropic dipole reorientation, because the corresponding phase-transition temperature should be higher than the decomposition temperature. Instead, quadrupolar order, i.e., the parallel orientation of the CN dumbbells without head-to-tail order, should prevail at high temperatures in AgCN. A comparison with the relaxation rates at the antiferroelectric transition temperatures in NaCN and KCN suggests a transition from this quadrupolar order to polar head-to-tail order at a temperature that is well consistent with $T_{c,p} \approx 475$ K, estimated from the relaxation strength. Based on these results, we propose the existence of a critical relaxation rate at the onset of dipolar order in the cyanides of $10^5$ - $10^7$ Hz, systematically varying with the cation radius.

Although the detected relaxation behavior in AgCN is supposed to occur below the polar ordering transition and, thus, should reflect the dynamics of a fraction of "free" dipoles that are not yet part of the polar order, it exhibits some characteristics of glassy freezing: The considerable broadening of the relaxation observed in the spectra evidences significant non-exponentiality of the dipolar relaxation, typical for glassy dynamics.[57,58] An orientational glass-transition temperature $T_g^o \approx 195$ K can be formally derived from the usual $\tau(T_g) \approx 100$ s criterion. In agreement with other cyanides,[2,8] the temperature-dependent relaxation rate of AgCN does not show any deviation from Arrhenius behavior. This corresponds to "strong" behavior within the strong-fragile classification of glass formers introduced by Angell.[76] In general, materials where the orientational degrees of freedom reveal glassy freezing tend to be rather strong glass formers.[5,77] However, one should be aware that the freezing-in of the highly restricted 180° flips of the CN dipoles in AgCN differs in many respects from the glassy freezing of the $\alpha$ relaxation in structural glasses and plastic crystals involving cooperative isotropic motions. Indeed, the symmetric broadening and pure Arrhenius behavior of the relaxation process observed in AgCN reminds of a Johari-Goldstein $\beta$-relaxation,[78] in accord with Ref. 7 suggesting that the $\alpha$ relaxation does not evolve in CuCN.

Finally, we want to remark that, at $T_g^o$, significant relaxation dynamics should still exist [$\Delta\varepsilon(195$ K$) > 0$; see Fig. 3], i.e., there are still reorienting CN dumbbells, not participating in the polar order. Their freezing-in upon further cooling will prevent the approach of full polar order involving all CN dipoles, for $T \to 0$ K.

## AUTHOR DECLARATIONS

### Conflict of Interest

The authors have no conflicts to disclose.

### Author Contributions

**Peter Lunkenheimer:** Writing – review & editing (lead); Investigation (lead); Formal analysis (lead); Visualization (lead). **Alois Loidl:** Writing – original draft (lead); Funding acquisition (lead); Project administration (lead). **Gyan P. Johari:** Conceptualization (lead); Writing – review and editing (supporting).

## DATA AVAILABILITY

The data that support the findings of this study are available from the corresponding author upon reasonable request.